\documentclass{article}
\usepackage{spconf,graphicx}
\usepackage{algorithm}
\usepackage{textcomp}
\usepackage{xcolor}
\usepackage{subcaption}
\usepackage{amsmath}
\usepackage{bm}
\usepackage{amsmath, amssymb, amsfonts}
\usepackage{algpseudocode}
\usepackage{algorithm}
\usepackage{enumitem}

\algrenewcommand\algorithmicrequire{\textbf{Require:}}
\algrenewcommand\algorithmiccomment[1]{\hfill\(\triangleright\)~#1}

\title{Channel Prediction under Network Distribution Shift Using Continual Learning-based Loss Regularization}
\name{
\begin{tabular}{c}
Muhammad Ahmed Mohsin\textsuperscript{1}, 
Muhammad Umer\textsuperscript{1},
Ahsan Bilal\textsuperscript{2},
Muhammad Ibtsaam Qadir\textsuperscript{3},\\
Muhammad Ali Jamshed\textsuperscript{4},
Dean F. Hougen\textsuperscript{2},
John M. Cioffi\textsuperscript{1}
\end{tabular}
} 
\address{
 \textsuperscript{1} Dept. of Electrical Engineering, Stanford University, Stanford, CA, USA\\
 \textsuperscript{2} School of Computer Science, University of Oklahoma, Norman, OK, USA\\
 \textsuperscript{3} Weldon School of Biomedical Engineering, Purdue University, West Lafayette, IN, USA\\
 \textsuperscript{4} College of Science and Engineering, University of Glasgow, Glasgow, UK
}

\begin{document}
%
\maketitle
\begin{abstract}
Modern wireless networks face critical challenges when mobile users traverse heterogeneous network configurations with varying antenna layouts, carrier frequencies, and scattering statistics. Traditional predictors degrade under distribution shift, with NMSE rising by 37.5\% during cross-configuration handovers. This work addresses catastrophic forgetting in channel prediction by proposing a continual learning framework based on loss regularization. The approach augments standard training objectives with penalty terms that selectively preserve network parameters essential for previous configurations while enabling adaptation to new environments. Two prominent regularization strategies are investigated: Elastic Weight Consolidation (EWC) and Synaptic Intelligence (SI). Across 3GPP scenarios and multiple architectures, SI lowers the high-SNR NMSE floor by up to 1.8 dB ($\approx$32--34\%), while EWC achieves up to 1.4 dB ($\approx$17--28\%). Notably, standard EWC incurs $\mathcal{O}(MK)$ complexity (storing $M$ Fisher diagonal entries and corresponding parameter snapshots across $K$ tasks) unless consolidated, whereas SI maintains $\mathcal{O}(M)$ memory complexity (storing $M$ model parameters), independent of task sequence length, making it suitable for resource-constrained wireless infrastructure \footnote{This work was done in collaboration with Intel Corporation, Samsung Electronics, and Ericsson}. 
\end{abstract}
\section{Introduction}\label{sec:intro}
Emerging and forthcoming wireless networks rely on advanced physical-layer designs, such as massive multiple-input multiple-output (MIMO), to meet escalating demands for throughput and reliability. These technologies' efficacy fundamentally depends on accurate and timely channel state information (CSI). The challenge grows with 5G New Radio (NR) specifications, which mandate uplink sounding intervals of at least 2 ms in time-division duplex (TDD) mode, while the channel coherence time at 28 GHz for a mobile user can be as short as 0.3 ms~\cite{villena2024aging}. This discrepancy necessitates a shift from mere channel estimation to proactive channel prediction, where future channel states are forecast to offset inherent CSI staleness $\rho(\Delta t) = J_0\!\big(2\pi f_D \Delta t\big)$, where $J_0(\cdot)$ is the zeroth-order Bessel function, $f_D$ is the maximum Doppler shift, and $\Delta t$ is the sounding delay. Failure to do so can result in substantial throughput degradation, with studies indicating that a 4 ms feedback delay at moderate mobility can reduce system sum-rate by approximately 50\%~\cite{li2021impact, umer2025neuralgaussianradiofields, 11011078}.

Learning-based predictors capture the non-linear dynamics of time-varying channels and often outperform statistical models~\cite{jiang2019neural}. However, when users traverse heterogeneous cells, distribution shifts in delay-spread, path-loss, and spatial-correlation can inflate NMSE by 37.5\% (1.8 dB)~\cite{mohsin2025continual}. By domain adaptation theory, $\epsilon_T(f) \leq \epsilon_S(f) + d_{\mathcal{H}}(S,T) + \lambda$, where growing source–target divergence $d_{\mathcal{H}}(S,T)$ under handovers explains this inflation and motivates continual learning.

A straightforward approach to handling this distribution shift is to fine-tune the model on new data, but this method leads to ``catastrophic forgetting,'' where a neural network abruptly loses previously acquired knowledge upon learning a new task~\cite{kirkpatrick2017overcoming}. For users revisiting prior cells, forgetting is non-optimal. Replay-based continual learning mitigates this but requires $\mathcal{O}(NK)$ memory, making it unsuitable for resource-constrained infrastructure. Thus, robust channel prediction is naturally a continual learning task~\cite{de2021continual}.

\begin{figure}[t]
    \centering
    \includegraphics[width=\linewidth]{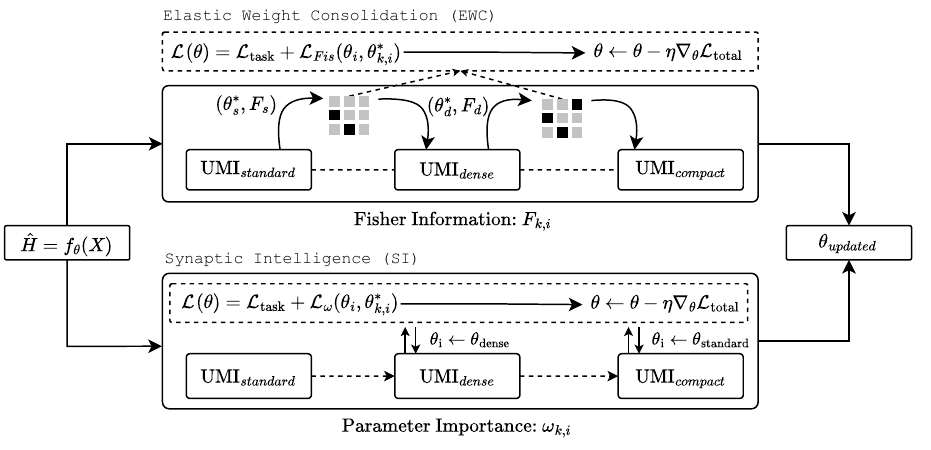}
    \caption{Comparison of EWC and SI for continual channel prediction. The framework illustrates task loss regularization with Fisher information ($F_{k,i}$) and parameter importance ($\omega_{k,i}$), enabling knowledge retention while adapting to distribution shifts.}
    \label{fig:ewc_si}
\end{figure}

This work addresses the challenge of catastrophic forgetting in channel prediction via loss regularization: augmenting the task loss with penalties that preserve parameters crucial for previous environments. The approach balances plasticity (learning new conditions) with stability (retaining prior knowledge). We evaluate two key strategies, Elastic Weight Consolidation (EWC) and Synaptic Intelligence (SI)~\cite{zenke2017continual}, as outlined in Algorithm~\ref{alg:unified} and Fig.~\ref{fig:ewc_si}.

\noindent\textbf{Contributions:} We frame channel prediction across heterogeneous network configurations as a continual learning task and propose a loss regularization framework that compares EWC and SI. 
We explicitly connect EWC’s quadratic penalty to max posterior estimate with a diagonal Gaussian prior (precision $F_{k,i}$, Equations~\ref{eq:fisher_expectation} and~\ref{eq:ewc_loss_fomrula}), while SI yields a data-driven precision via online weight accumulation as in Equation~\ref{eq:si_weight_accumulation}. 
This clarifies why SI is more stable under mini-batch noise and non-stationarity typical of mobile channels. 
Experiments show that EWC yields 0.8--1.4 dB ($\approx$17--28\%) NMSE reduction while SI achieves up to 1.8 dB ($\approx$32--34\%), with task-independent $O(M)$ memory, making the approach viable across diverse deep-learning backbones and 3GPP scenarios.

\begin{algorithm}[tb]
\caption{Unified Loss-Regularized Continual Channel Prediction (EWC / SI)}
\label{alg:unified}
\small
\begin{algorithmic}[1]
\Require Learning rate $\eta$; regimen $r\!\in\!\{\mathrm{EWC},\mathrm{SI}\}$; EWC coeff. $\alpha$; SI coeff. $\beta$; damping $\xi$
\State Initialize parameters $\theta$;\; \textbf{if} $r=\mathrm{EWC}$ \textbf{then} $\mathcal{B}\gets\emptyset$ \textbf{else} $\omega_i,\tilde{\omega}_i\gets0,\;\theta_i^{\mathrm{ref}}\gets\theta_i\ \forall i$
\For{each task $D_k$}
  \For{each mini-batch $(\mathbf{X},\mathbf{H})\in D_k$}
    \State $\ell_{\mathrm{task}}\!\gets\!\ell_{\mathrm{NMSE}}\!\big(\mathbf{H},f_{\theta}(\mathbf{X})\big)$
    \State $\ell_{\mathrm{reg}}\!\gets\!\begin{cases}
      \frac{\alpha}{2}\!\!\sum_{(\theta_j^{*},F_j)\in\mathcal{B}}\sum_i F_{j,i}\!\big(\theta_i-\theta^{*}_{j,i}\big)^{2}, & r=\mathrm{EWC} \\
      \frac{\beta}{2}\sum_i \omega_i\!\big(\theta_i-\theta_i^{\mathrm{ref}}\big)^{2}, & r=\mathrm{SI}
    \end{cases}$
    \State $\theta \gets \theta - \eta\,\nabla_{\theta}\!\big(\ell_{\mathrm{task}}+\ell_{\mathrm{reg}}\big)$
    \If{$r=\mathrm{SI}$}
      \State $\tilde{\omega}_i \gets \tilde{\omega}_i + \big(\partial_{\theta_i}\ell_{\mathrm{task}}\big)^{2}\eta \ \ \forall i$
    \EndIf
  \EndFor
  \If{$r=\mathrm{EWC}$}
    \State $F_{k,i}\gets \frac{1}{|D_k|}\!\sum_{(\mathbf{X},\mathbf{H})\in D_k}\!\big(\partial_{\theta_i}\ell_{\mathrm{NMSE}}\big)^{2}$;\quad
           $\mathcal{B}\gets \mathcal{B}\cup\{(\theta_k^{*}\!\gets\!\theta,\,F_k)\}$
  \Else
    \State $\Delta\theta_i\gets \theta_i-\theta_i^{\mathrm{ref}}$;\;
           $\omega_i\gets \omega_i + \frac{\tilde{\omega}_i}{(\Delta\theta_i)^{2}+\xi}$;\;
           $\tilde{\omega}_i\gets 0$;\; $\theta_i^{\mathrm{ref}}\gets\theta_i$
  \EndIf
\EndFor
\end{algorithmic}
\end{algorithm}
\section{Problem Formulation}\label{sec: prob_form}
A sequence of channel prediction tasks $\mathcal{T} = \{\mathcal{T}_1, \mathcal{T}_2, \ldots, \mathcal{T}_K\}$ is considered, where each task $\mathcal{T}_k$ corresponds to a specific network configuration with dataset $\mathcal{D}_k = \{(\mathbf{X}_i^{(k)}, \mathbf{H}_i^{(k)})\}_{i=1}^{N_k}$. Here, $\mathbf{X}_i^{(k)} \in \mathbb{C}^{2 \times T \times N_{\text{tx}} \times N_{\text{rx}}}$ represents a sequence of $T$ past channel realizations (with real and imaginary parts separated), and $\mathbf{H}_i^{(k)} \in \mathbb{C}^{2 \times N_{\text{tx}} \times N_{\text{rx}}}$ is the target channel matrix to be predicted.

When a neural network with parameters $\bm{\theta}$ is sequentially trained on tasks $\mathcal{T}_1, \mathcal{T}_2, \ldots, \mathcal{T}_k$, naive fine-tuning leads to catastrophic forgetting~\cite{varshney2023deep, olickal2023lstm}, where the model's performance on earlier tasks $\mathcal{T}_1, \ldots, \mathcal{T}_{k-1}$ severely degrades. Empirical studies indicate 37.5\% increase in normalized NMSE  (as discussed in Section~\ref{sec:results_evaluation}) when transitioning between network configurations.

Let $f_{\bm{\theta}}: \mathbb{C}^{2 \times T \times N_{\text{tx}} \times N_{\text{rx}}} \rightarrow \mathbb{C}^{2 \times N_{\text{tx}} \times N_{\text{rx}}}$ denote the channel prediction function parameterized by $\bm{\theta}$. The prediction loss for task $k$ is defined as:
\begin{equation}
\mathcal{L}_k(\bm{\theta}) = \frac{1}{|\mathcal{D}_k|} \sum_{(\mathbf{X}, \mathbf{H}) \in \mathcal{D}_k} \frac{\|\mathbf{H} - f_{\bm{\theta}}(\mathbf{X})\|_F^2}{\|\mathbf{H}\|_F^2},
\end{equation}
where $\|\cdot\|_F$ denotes the Frobenius norm. The challenge is to minimize the current task loss $\mathcal{L}_k(\bm{\theta})$ while preserving performance on previous tasks, i.e., avoiding significant increases in $\mathcal{L}_j(\bm{\theta})$ for $j < k$.

\begin{figure*}[t!]
     \centering
    \begin{subfigure}[t]{0.31\textwidth} 
         \centering
         \includegraphics[width=\textwidth]{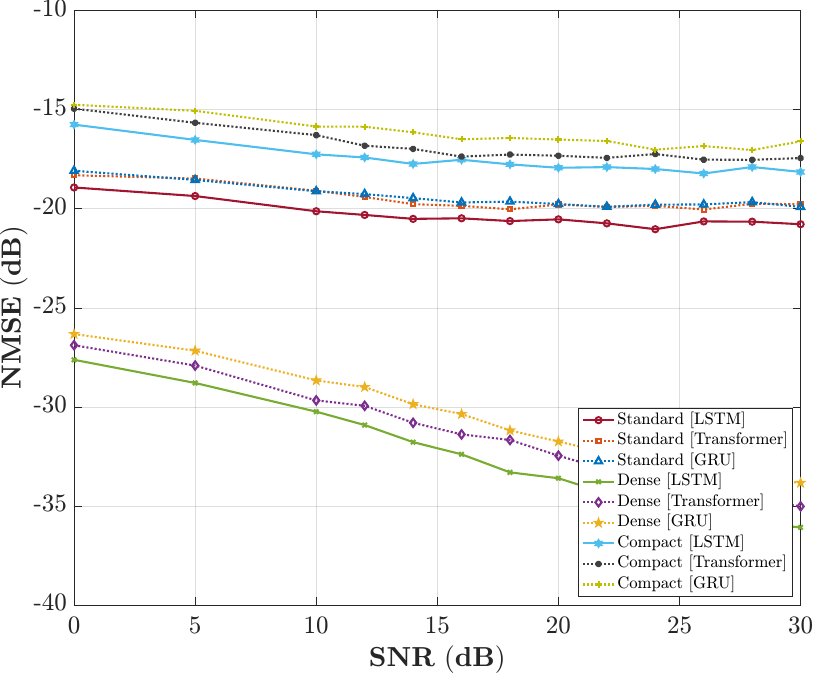}
         \caption{SNR vs. NMSE across LSTM/GRU/Transformer backbones}
         \label{fig:EpochsvNMSE}
     \end{subfigure}
     \hspace{0.2em} 
     \begin{subfigure}[t]{0.31\textwidth} 
         \centering
         \includegraphics[width=\textwidth]{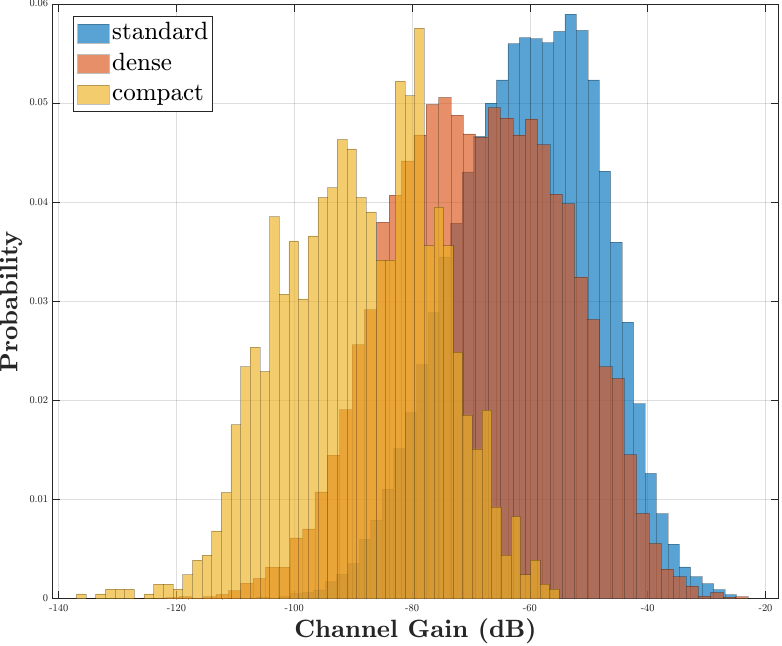}
         \caption{Channel-gain distributions for UMi compact/dense/standard}
         \label{fig:EpochsvLoss}
     \end{subfigure}
     \hspace{0.2em} 
     \begin{subfigure}[t]{0.31\textwidth} 
         \centering
         \includegraphics[width=\textwidth]{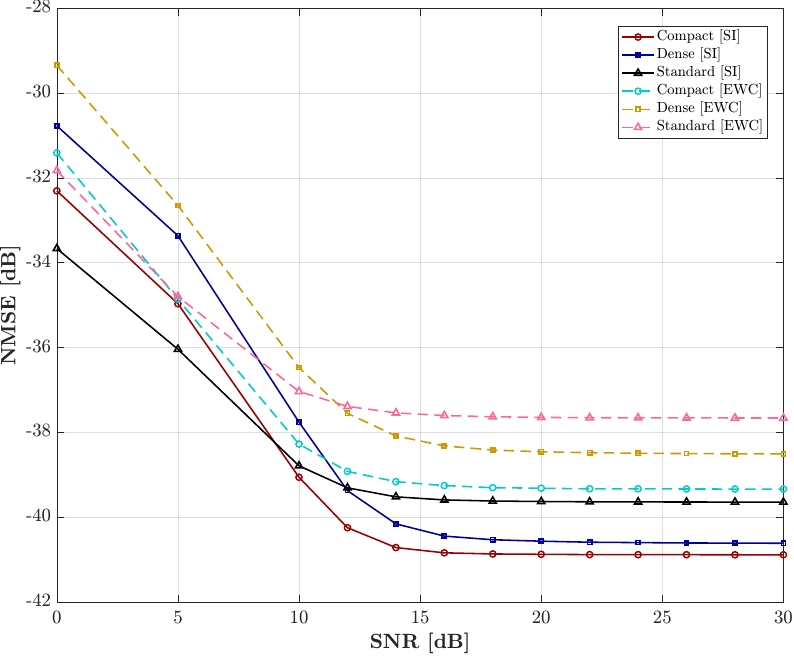} 
         \caption{SI vs. EWC (SNR (dB) vs. NMSE (dB))}
         \label{fig:SNRvNMSE}
     \end{subfigure}
     \caption{EWC vs. SI for continual channel prediction}
     \label{fig:2}
\end{figure*}

Loss regularization methods address catastrophic forgetting by augmenting the training objective with penalty terms that selectively constrain parameter updates based on their importance to previous tasks. Unlike replay-based approaches that require $O(NK)$ memory for storing $N$ samples across $K$ tasks, regularization techniques maintain $O(M)$ memory complexity for $M$ model parameters, enabling deployment on resource-constrained wireless infrastructure. The fundamental insight leverages the heterogeneous contribution of network parameters to prediction accuracy across different propagation environments. Some parameters encode task-agnostic temporal dependencies, while others capture configuration-specific channel statistics that must be preserved during network transitions. By formulating continual learning as a constrained optimization problem where parameter importance weights $\omega_i$ modulate regularization strength (as shown in Fig.~\ref{fig:2}), these methods preserve knowledge without hindering adaptation~\cite{chen2022continual}, with constant overhead independent of network configurations.

\section{Methodology}
\label{sec:methodology}
This work proposes a loss regularization framework for continual channel prediction that integrates two complementary approaches: EWC and SI, as in Fig.~\ref{fig:ewc_si}. Both methods augment the standard channel prediction loss with regularization terms that preserve knowledge of previous network configurations~\cite{qu2025recent}. The loss regularization framework modifies the standard training objective by adding penalty terms that discourage changes to parameters important for previous tasks. The general form of the regularized loss function is:
\begin{equation}
\mathcal{L}_{\text{total}}(\bm{\theta}) = \mathcal{L}_{\text{task}}(\bm{\theta}) + \mathcal{L}_{\text{reg}}(\bm{\theta}),
\end{equation}
where $\mathcal{L}_{\text{task}}(\bm{\theta})$ is the standard NMSE loss for the current task, and $\mathcal{L}_{\text{reg}}(\bm{\theta})$ is the regularization term that preserves knowledge from previous tasks.\\
\textbf{EWC} identifies critical parameters by estimating their importance using the Fisher Information Matrix (FIM) and penalizes deviations from the optimal parameter values found for previous tasks~\cite{kirkpatrick2017overcoming, huszar2018note}. For each completed task $\mathcal{T}_k$ with optimal parameters $\bm{\theta}_k^*$, 
the diagonal FIM $\mathbf{F}_k$ is empirically computed as:
\begin{equation}
F_{k,i} = \frac{1}{|\mathcal{D}_k|} 
\sum_{(\mathbf{X}, \mathbf{H}) \in \mathcal{D}_k} 
\left(\frac{\partial \mathcal{L}_{\text{NMSE}}(\bm{\theta}_k^*; \mathbf{X}, \mathbf{H})}
{\partial \theta_i}\right)^2 .
\label{eq:fisher_nmse}
\end{equation}
This NMSE-based approximation can be formulated as likelihood-based Fisher information:
\begin{equation}
F_{k,i} \approx \mathbb{E}_{(\mathbf{X},\mathbf{H}) \sim \mathcal{D}_k}
\left[ \left( \frac{\partial}{\partial \theta_i} 
\log p(\mathbf{H}\mid \mathbf{X}; \bm{\theta}_k^*) \right)^2 \right].
\label{eq:fisher_expectation}
\end{equation}
Parameters with high Fisher information are considered critical for maintaining performance on that task. The EWC regularization term for a single previous task is:
\begin{equation}
\label{eq:ewc_loss_fomrula}
\mathcal{L}_{\text{EWC}}^{(k)}(\bm{\theta}) = \frac{\alpha}{2} \sum_{i} F_{k,i} (\theta_i - \theta_{k,i}^*)^2,
\end{equation}
where $\alpha > 0$ is a stability coefficient that controls the strength of regularization and $\theta_{k,i}^*$ denotes the value of parameter $\theta_i$ after training on task $\mathcal{T}_k$. 
This quadratic penalty can be interpreted as Maximum a Posteriori (MAP) estimation, where the first term corresponds to the negative log-likelihood of the current task data and the second term encodes a Gaussian prior with precision $F_{k,i}$:
\begin{equation}
\hat{\bm{\theta}}_{\text{MAP}} 
= \arg\min_{\bm{\theta}} \Big[ -\log p(\mathcal{D}_k \mid \bm{\theta}) 
+ \tfrac{\alpha}{2}\sum_i F_{k,i}(\theta_i - \theta_{k,i}^*)^2 \Big].
\end{equation}
Here, $F_{k,i}$ is the Fisher information (Equation.~\ref{eq:fisher_expectation}), linking the curvature of the log-likelihood to the strength of the Gaussian prior. For multiple previous tasks, the EWC loss becomes:
\begin{equation}
\mathcal{L}_{\text{EWC}}(\bm{\theta}) = \frac{\alpha}{2} \sum_{i} \sum_{j=1}^{k} F_{j,i} (\theta_i - \theta_{j,i}^*)^2 .
\end{equation}
The EWC algorithm first initializes model parameters and a parameter bank. For each network configuration, it processes mini-batches using gradient descent on the combined task and regularization loss. Subsequently, it computes Fisher information matrices and stores optimal parameters with their importance weights for future regularization.\\
\textbf{SI} addresses EWC's computational and memory limitations by tracking parameter importance online during training, without requiring explicit Fisher information computation~\cite{zenke2017continual, puiu2022rethinking, van2025computation}. During training on task $\mathcal{D}_k$, SI maintains running estimates of parameter importance $\tilde{\omega}_i$ for each parameter $\theta_i$:
\begin{equation}
\tilde{\omega}_i \leftarrow \tilde{\omega}_i + \left(\frac{\partial \mathcal{L}_{\text{NMSE}}}{\partial \theta_i}\right)^2 \eta ,
\label{eq:si_running}
\end{equation}
where $\eta$ is the learning rate. This accumulation captures the total contribution of each parameter to loss reduction during the current task. Equivalently, this update can be interpreted as a path integral of the work performed by parameter $i$ along the optimization trajectory. Let $g_i^t = \partial_{\theta_i}\mathcal{L}_{\text{NMSE}}(\theta^t)$ and $\Delta\theta_i^t = \theta_i^{t+1}-\theta_i^t$. Then, the total work is
\begin{equation}
\mathcal{W}_i^{(k)} \;\approx\; \sum_{t\in \mathcal{T}_k} g_i^t \,\Delta\theta_i^t \;\approx\; \sum_{t\in \mathcal{T}_k} (g_i^t)^2 \eta,
\label{eq:si_work}
\end{equation}
which shows that $\tilde{\omega}_i$ in Eq.~\eqref{eq:si_running} is a discrete approximation of the optimization path integral. After completing task $k$, SI normalizes this accumulated work by the squared displacement of each parameter:
\begin{equation} 
\label{eq:si_weight_accumulation} 
\omega_i \leftarrow \omega_i + \frac{\tilde{\omega}_i}{(\Delta\theta_i)^2 + \xi} , 
\end{equation}
where $\Delta\theta_i = \theta_i - \theta_i^{(0)}$ is the total change in parameter $i$ during task $k$, and $\xi > 0$ is a small damping term to prevent division by zero. This yields a final importance score $\omega_i$ that captures how much ``useful work'' each parameter performed relative to its displacement. The SI regularization term is then:
\begin{equation}
\mathcal{L}_{\text{SI}}(\bm{\theta}) = \frac{\beta}{2} \sum_{i} \omega_i \big(\theta_i - \theta_i^{(0)}\big)^2 ,
\label{eq:si_reg}
\end{equation}
where $\beta > 0$ is a stability coefficient analogous to $\alpha$ in EWC, and $\theta_i^{(0)}$ are the reference parameters at the start of the task. Training proceeds by gradient descent on the combined objective:
\begin{equation}
\theta_i^{t+1} = \theta_i^{t} - \eta \left( 
\frac{\partial \mathcal{L}_{\text{task}}}{\partial \theta_i} +
\lambda \frac{\partial \mathcal{L}_{\text{SI}}}{\partial \theta_i}
\right),
\label{eq:si_update}
\end{equation}
where $\lambda=\beta$ scales the SI penalty. This unifies SI with other loss-regularization methods as a penalized optimization process constraining updates along critical parameter directions.

EWC and SI present complementary trade-offs: EWC requires $O(MK)$ memory and doubles training time through Fisher information computation, while SI maintains $O(M)$ memory with negligible computational overhead via online importance tracking. The stability–plasticity trade-off can be expressed as a constrained optimization:
\begin{equation}
\min_{\bm{\theta}} \; \mathbb{E}_{\mathcal{T}_k}[\mathcal{L}_{\text{task}}(\bm{\theta})]
\quad \text{s.t.} \quad
\Delta \mathcal{L}_{\text{prev}} \leq \epsilon,
\end{equation}
where $\Delta \mathcal{L}_{\text{prev}}$ is the increase in loss on previous tasks. In practice, this constraint is relaxed via quadratic penalties (EWC) or online importance reweighting (SI), enforcing a maximum allowable loss increase $\epsilon$ on past tasks.

SI is more robust, as it accumulates parameter importance throughout training, whereas EWC's Fisher approximation can be unreliable under high gradient stochasticity. The framework integrates with Long Short-Term Memory (LSTM), Gated Recurrent Unit (GRU), and Transformers via parameter regularization: $\alpha,\beta\in[0.1,1.0]$ trade off plasticity–stability and $\xi\in[10^{-6},10^{-3}]$ stabilizes SI. Ranges come from a grid search on held-out Urban Microcell (UMi) validation across backbones, with $\xi$ preventing near-zero denominators in $(\Delta\theta)^2+\xi$.

\section{Performance Evaluation} \label{sec:results_evaluation}
Experimental evaluation demonstrates the effectiveness of the proposed loss regularization framework across three representative UMi scenarios with varying propagation complexity, utilizing LSTM, GRU, and Transformer architectures for channel prediction with constrained replay memory as in Fig.~\ref{fig:2}. The comprehensive assessment shows that SI consistently surpasses EWC by leveraging online importance tracking, which provides more reliable parameter estimates under mini-batch stochasticity. In UMi--\emph{compact} scenarios, SI achieves \textbf{1.7--1.8 dB} NMSE improvements (\(\approx 34\%\) reduction), while in dense and standard settings, it maintains \textbf{0.8--1.4 dB} gains, as illustrated in Fig.~\ref{fig:SNRvNMSE}.

The proposed framework exhibits exceptional robustness in SNR versus NMSE characterization, maintaining sharp error reduction in critical 8-12 dB SNR regimes while avoiding saturation artifacts observed in conventional methods at high SNRs (Fig.~\ref{fig:SNRvNMSE}). Performance analysis at 20 dB SNR reveals SI's superior capability to maintain low error floors across varying channel conditions, a critical requirement for practical wireless infrastructure deployment. Online accumulation of parameter importance yields more accurate preservation of channel prediction across environments than post-hoc FIM.

Architecture-agnostic performance analysis reveals consistent improvements across LSTM, Transformer, and GRU implementations, with LSTM architectures delivering optimal performance for both regularization approaches, as shown in Fig.~\ref{fig:EpochsvNMSE}. From compact to standard scenarios, prediction complexity increases, yet SI sustains smaller performance gaps than baselines, with consistent 0.8–1.4 dB gains across dense and standard UMi settings (Fig.~\ref{fig:SNRvNMSE}). With $O(M)$ memory and negligible overhead, SI preserves prior knowledge under distribution shifts, enabling reliable continual channel prediction without replay or task-specific storage.

\section{Conclusion and Future work} \label{sec: results}
Continual channel prediction hinges on balancing plasticity and stability. Naive fine-tuning disrupts this balance, while loss regularization preserves critical parameters for stable adaptation. SI’s online importance accumulation proves more reliable than Fisher-based EWC, underscoring that memory encoding design is as vital as model capacity for robust continual learning in wireless systems.

Future work includes adaptive regularization that tunes the stability–plasticity trade-off under mobility and coherence-time variations, extending the framework to massive MIMO, multi-cell cooperation, and federated continual learning for scalability. Further directions involve parameter-importance quantization in 3GPP CSI-feedback pipelines and validation in ultra-dense deployments with reconfigurable intelligent surfaces, while maintaining lightweight memory and computation.  
\bibliographystyle{IEEEbib}
\bibliography{main}

\end{document}